\documentclass{moriond}

\begin{document}
\vspace*{4cm}
\title{PROBING INFLATION MODELS WITH GRAVITATIONAL WAVES}

\author{ VALERIE DOMCKE }

\address{ AstroParticule et Cosmologie (APC)/Paris Centre for Cosmological Physics, Universit\'e Paris Diderot}

\maketitle

\begin{center}
\begin{minipage}{5.2truein}
                 \footnotesize
                 \parindent=0pt
A direct detection of primordial gravitational waves is the ultimate probe for any inflation model. 
While current CMB bounds predict the generic scale-invariant gravitational wave spectrum from slow-roll inflation to be below the reach of upcoming gravitational wave interferometers,  this prospect may dramatically change if the inflaton is a pseudoscalar. In this case, a coupling to any abelian gauge field leads to a tachyonic instability for the latter and hence to a new source of gravitational waves, directly related to the dynamics of inflation.
In this contribution we discuss how this setup enables the upcoming gravitational wave interferometers advanced LIGO/VIRGO and eLISA to probe the microphysics of inflation, distinguishing between different universality classes of single-field slow-roll inflation models. We find that the prime candidate for an early detection is a Starobinsky-like model. \par
\end{minipage}\end{center}
\vskip 2em \par

\section{Introduction}

Unlike any other messenger, gravitational waves (GWs) can travel freely through the Universe, carrying information on times as early as cosmic inflation.
During inflation, quantum fluctuations of the inflaton field and of the metric are stretched and become large-scale classical perturbations.
GWs, corresponding to the metric fluctuations, carry the imprint of this very early stage of our universe, with a 1:1 correspondence between their frequency, the scale of the corresponding perturbation mode and the point in time when this mode exited the horizon during inflation.
 The frequency-dependence of the GW spectrum thus directly encodes the dynamics of cosmic inflation, i.e.\ can give us a time-resolved view of the microphysics (eg.\ the scalar potential) responsible for inflation.

Current indirect bounds on primordial GWs, obtained from bounds on the Cosmic Microwave Background (CMB) polarization, indicate that the nearly scale-invariant spectrum of the vacuum fluctuations during slow-roll inflation lies below the range of current and upcoming direct GW detectors. This picture may however change dramatically once interactions of the inflaton with other particles are taken into account~\cite{Cook:2011hg}. 
To this end, we consider the generic coupling of a \emph{pseudoscalar} inflaton $\phi$ to the field strength tensor $F_{\mu \nu}$ of an abelian gauge group,
\begin{equation}
{\cal L}_\textrm{int} \sim \phi F_{\mu \nu} \tilde{F}^{\mu \nu} \,.
\label{eq:Lint}
\end{equation}
Such pseudoscalar  (or `axionic') flat directions, suitable for inflation, may be expected to be abundant at the high energy scales of cosmic inflation (for supergravity embeddings, see eg.~\cite{Kawasaki:2000yn,sugraexamples}). 
As we will see, the presence of the interaction term~(\ref{eq:Lint}) leads to a non-perturbative production of the gauge field during inflation~\cite{pproduction}, which provides an additional source of tensor perturbations, leading to a large enhancement of the resulting GW signal at small scales. In this contribution, we discuss how this enhancement will enable GW interferometers to probe different universality classes of inflation~\cite{Mukhanov:2013tua}, classified by properties of the underlying scalar potential $V(\phi)$. This proceeding is based on work with Pierre Bin\'etruy and Mauro Pieroni~\cite{Domcke:2016bkh}.

\section{Pseudoscalar inflation}

We consider a pseudoscalar $\phi$ coupled to ${\cal N}$ abelian gauge fields $A_a^\mu$ (see also \cite{axioninflation} and refs therein),
\begin{equation}
{\cal L}=  -\frac{1}{2} \partial_\mu \phi \partial^\mu \phi  - \frac{1}{4} F^a_{\mu \nu} F_a^{\mu \nu} - V(\phi) - \frac{\alpha^a}{4 \Lambda} \phi F^a_{\mu \nu} \tilde F_a^{\mu \nu} \,,
\end{equation}
with $F^a_{\mu \nu}$ ($\tilde F_a^{\mu \nu}$) denoting the (dual) field-strength tensor, $\alpha/\Lambda$ parameterizing the coupling between the pseudoscalar and the gauge fields and $V(\phi)$ denoting the scalar potential driving inflation. The resulting equations of motion for the classical background fields read
\begin{equation}
\label{eq:eq_motion}
\ddot \phi + 3 H \dot \phi + \frac{\partial V}{\partial \phi} = \frac{\alpha}{\Lambda} \langle \vec E \vec B \rangle \,, \qquad
\frac{ d^2 A^a_\pm(\tau, k)}{d \tau^2} + \left( k^2 \mp \frac{2 \xi k }{ \tau } \right)A^a_\pm(\tau, k) = 0 \,, 
\end{equation}
with
$\xi \equiv \alpha |\dot \phi| /(2 \Lambda H)$.
Here the second equation refers to the Fourier modes of the gauge field, $\vec{A}^{a} =\vec{e}_{\pm} A_{\pm}^a \exp(i \vec{k}\vec{x})$. The subscript $\pm$ denotes the two helicity modes of the massless gauge field. One of them (in our notation the $A_+$ mode) experiences a tachyonic instability, leading to an exponential growth of the low-momentum modes. If $\xi$ is a slowly varying function in time, the two equations  in (\ref{eq:eq_motion})  decouple and we can express the classical background gauge field as
\begin{equation}
A_+^a \simeq \frac{1}{\sqrt{2k}} \left( \frac{k}{2 \xi a H}\right)^{1/4} e^{ \pi \xi - 2 \sqrt{2 \xi k/(a H)}} \,.
\end{equation}
With this we find
$\langle \vec E \vec B \rangle \simeq {\cal N} \cdot \,  2.4 \cdot 10^{-4} H^4/\xi^4 e^{2 \pi \xi} $. 
This acts an additional friction term in the slow-roll equation of motion for $\phi$, which will overcome the Hubble friction term for sufficiently large values of $\xi$, thus modifying the background dynamics of inflation (and thus the predictions of a given inflation model, e.g.\ $n_s$ and $r$). Note that $\xi \sim \dot \phi/H \sim \sqrt{\epsilon}$, with $\epsilon$ denoting the first slow-roll parameter of single-field inflation. In most single-field inflation models, $\epsilon$ is very small when the CMB scales exited the horizon (constrained by the CMB data) but increases monotonously to reach $\epsilon = 1$ at the end of inflation. Hence all effects sourced by the gauge fields are expected to be strongly suppressed at the CMB scales but can be very large at smaller scales, corresponding to perturbations mode which exited the horizon towards the end of inflation.

The gauge field background further modifies the equation of motion for the fluctuations of $\phi(t)$, 
leading to an additional contribution to the scalar power spectrum~\cite{Linde:2012bt},
\begin{equation}
\Delta_s^2(k) = \Delta_s^2(k)_\textrm{vac} + \Delta_s^2(k)_\textrm{gauge} = \left(\frac{H^2}{2 \pi |\dot \phi|}\right)^2 + \left( \frac{\alpha \langle \vec E \vec B \rangle/ \sqrt{\cal N}}{3 \beta \Lambda H \dot \phi} \right)^2 ,
\label{eq:scalar}
\end{equation}
with
$\beta \equiv 1 - 2 \pi \xi \alpha \langle \vec E \vec B \rangle /(3 \Lambda H \dot \phi)$. At large values of $\xi$, this leads to an approximately scale invariant spectrum, $\Delta_s^2 \simeq 1/(4 \pi^2 {\cal N} \xi^2)$.

Similarly, the gauge field background also provides an additional source for the tensor fluctuations, leading to an additional contribution to the GW spectrum
\begin{equation}
\Omega_{GW} = \frac{1}{12} \Omega_{R,0} \left( \frac{H}{ \pi M_P} \right)^2 (1 + 4.3 \cdot 10^{-7} {\cal N} \frac{H^2}{M_P^2 \xi^6} e^{4 \pi \xi})\,,
\label{eq:OmegaGW}
\end{equation}
with $\Omega_{R,0} = 8.6 \cdot 10^{-5}$ denoting the radiation energy density today and $M_P = 2.4 \cdot 10^{18}$~GeV denoting the reduced Planck mass. Remarkably, the new contribution to the GW spectrum is maximally chiral, providing a powerful lever to distinguish it from other stochastic GW backgrounds.

\section{Probing the microphysics of inflation}

The impact of the gauge fields depends sensitively on the parameter $\xi$. 
As long as the gauge fields are subdominant in the equation of motion for $\phi$,  we have $ \xi^2 \sim \epsilon = \dot \phi/(2 H^2) \simeq \epsilon_V = (V'/V)^2/(2 M_P^2)$, i.e.\ the parameter $\xi$ directly probes the scalar potential $V(\phi)$. Exploiting that the vast majority of single-field slow-roll inflation models may be parametrized by~\cite{Mukhanov:2013tua}
\begin{equation}
\label{eq:Nparameterization}
\epsilon_V \simeq  \frac{\beta}{N^p}  + {\cal O}(1/N^{p+1})\,,
\end{equation}
with $N$ denoting the number of e-folds ($N = -\int H dt$) elapsed since the end of inflation, we see that the predictions for pseudoscalar inflation can be phrased in terms of only three parameters: $\alpha/\Lambda$, $\beta$ and $p$.  This allows us to systematically study the parameter space and to identify how future measurements may distinguish between different universality classes of inflation (different values of $p$) in this context.

\begin{figure}[t]
\centering
\begin{minipage}{0.45 \textwidth}
{\includegraphics[width=\columnwidth]{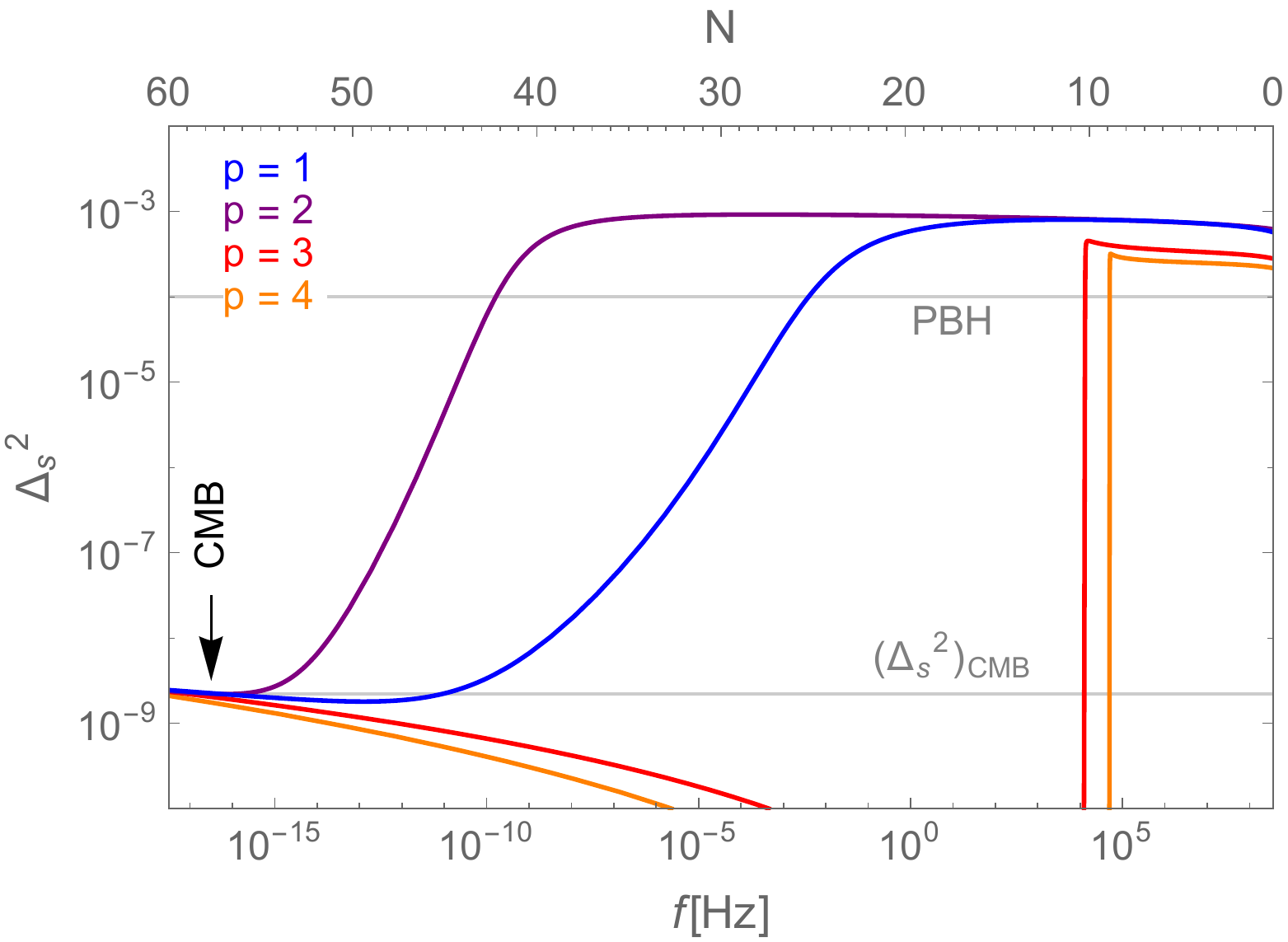}} 
\end{minipage}
\qquad
\begin{minipage}{0.45 \textwidth}
{\includegraphics[width=\columnwidth]{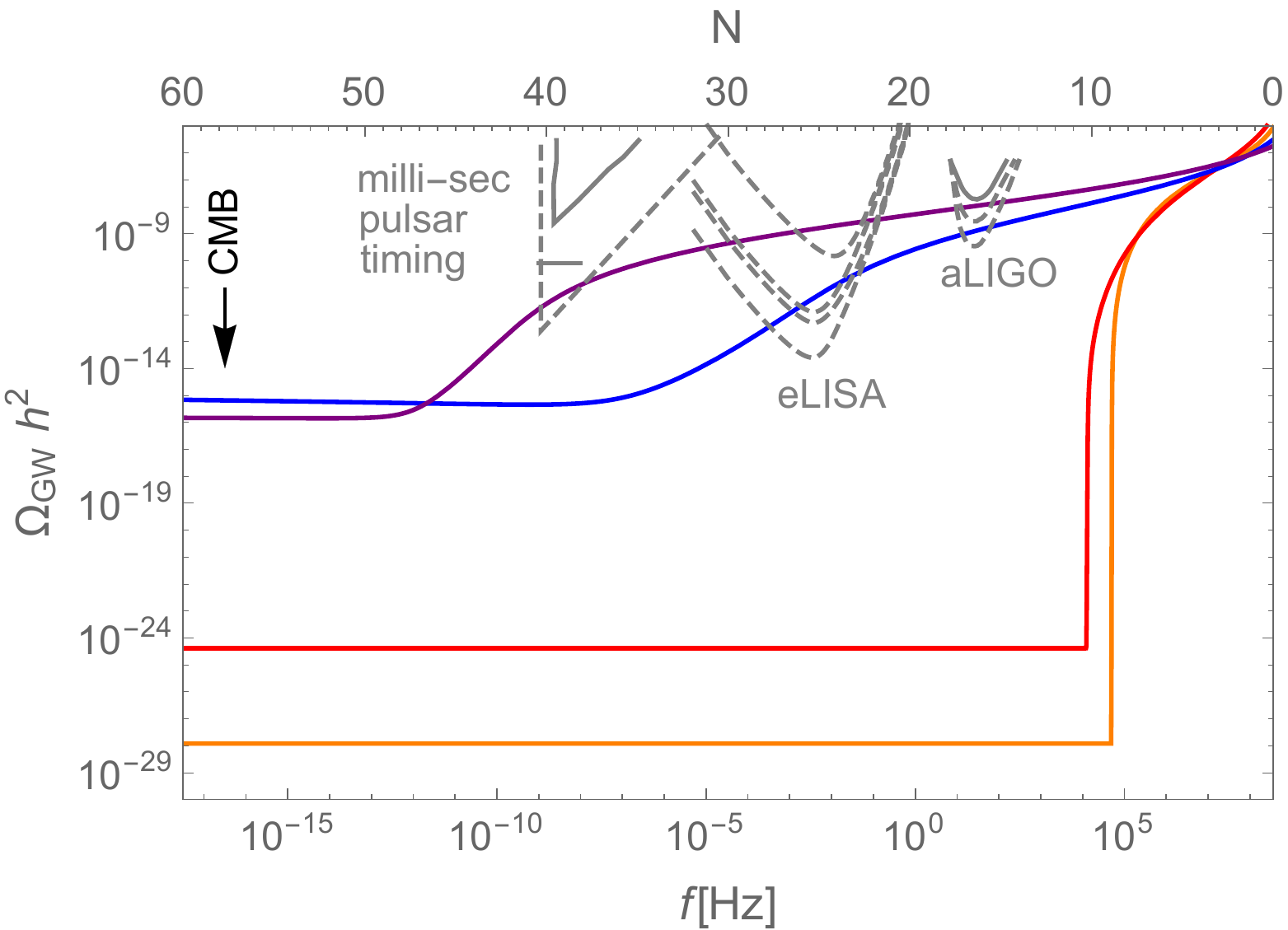}} 
\end{minipage}
\caption{Scalar and tensor power spectrum for different universality classes of inflation. For the scalar power spectrum, we show the CMB constraint as well as the PBH bound. For the tensor power spectrum, we indicate the sensitivity of some current and upcoming GW experiments.}
\label{fig:powerspectra}
\end{figure}

Fig.~\ref{fig:powerspectra} illustrates this for the scalar and tensor power spectrum, using four exemplary inflation models: quadratic chaotic inflation ($p = 1$), Starobinsky inflation ($p = 2$) and two implementations of hilltop inflation ($p = \{3,4\}$)~\cite{inflationmodels}. The two parameters $\beta$ and $\alpha/\Lambda$ have been chosen to maximize the GW spectrum while obeying the CMB constraints.
As can be easily understood from the $\xi$-dependence of the expressions above, a larger value of the parameter $p$ implies suppression of the amplitude of the tensor fluctuations at CMB scales, but also a steeper increase of the spectrum towards higher frequencies. The parameter $\alpha/\Lambda$ governs the onset of the strong gauge field regime, i.e.\ increasing $\alpha/\Lambda$ shifts the strong increase in the spectrum to lower frequencies.\footnote{The late rise in the hilltop models in Fig.~\ref{fig:powerspectra} is due to CMB constraints on the spectral index $n_s$, which force $\alpha/\Lambda$ to be small in these models. Since $n_s \simeq 1 - p/N$ for $p > 1$, models with large value of $p$ feature low values of $n_s$ compared to the current Planck data, a situation which is aggravated in the presence of gauge fields due to the extra friction term.} Note that both the scalar and tensor spectrum approach nearly universal values in the strong gauge field regime.\footnote{Both spectra obtain a $1/{\cal N}$ suppression in this regime if there is more than one abelian gauge group present. It should be noted that this regime of strong gauge fields is subject to a number of theoretical uncertainties.}

From Fig.~\ref{fig:powerspectra} we see that from the point of view of potential observations, the Starobinsky-type model with $p = 2$ is the most promising candidate. We thus turn to a more detailed study of the parameter space of this model, cf.\ Fig.~2.
\begin{figure}[t]
\centering
\begin{minipage}{0.4 \textwidth}
\includegraphics[width=1\columnwidth]{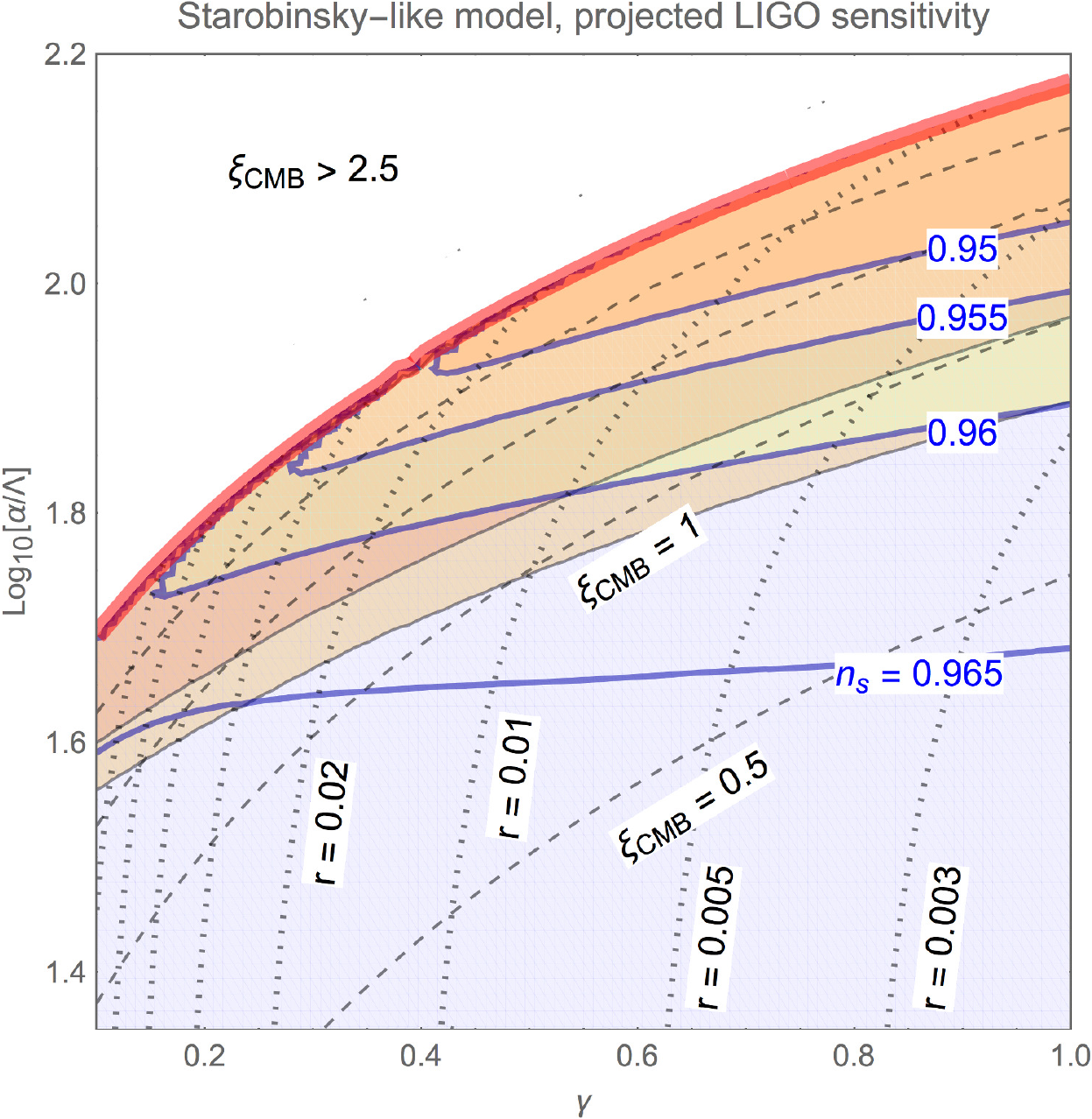}
\label{fig:scan_plots}
\end{minipage} \hfil
\begin{minipage}{0.5 \textwidth}
\footnotesize{

\vspace{-0.5cm}

Figure 2 - Plot of the $(\alpha/\Lambda, \gamma)$ parameter space for the Starobinsky model,
\begin{equation}
V(\phi) = V_0 \, (1 -  e^{- \gamma \phi})^2\,,
\end{equation}
($p = 2, \, \beta = 1/(2 \gamma^2)$) with contour lines for the spectral index $n_s$ (solid blue), the tensor-to-scalar ratio $r = \{0.003, 0.005, 0.1, 0.2, 0.3, \dots \}$ (dotted) and $\xi_\textrm{CMB} = \{ 0.5, 1, 1.5, \dots \}$ (dashed). The white region in the top left corner is excluded due to the non-observation of non-gaussianities in the CMB. The orange shaded regions denote the projected sensitivity for advanced LIGO in the O2 and O5 run. The planned space-based interferometer eLISA will probe a similar region of the parameter space as LIGO, demonstrating the power of future multi-frequency GW analysis. See~\cite{Domcke:2016bkh} for details.}
\end{minipage}
\end{figure}
 Here we in particular point out the powerful complementarity between CMB measurements  and GW interferometers. 
Further important predictions of this setup are the production of primordial black holes~\cite{Linde:2012bt}, the predictions of excess radiation in form of gravitational waves contributing to the measurement of $N_\textrm{eff}$
as well as the prediction of possibly observable excess $\mu$-distortions in the CMB black body spectrum~\cite{Meerburg:2015zua}.

\section{Conclusion}

If the inflaton is a pseudoscalar, its coupling to any massless abelian gauge field leads to a non-perturbative production of the latter during inflation. This in return modifies the inflationary dynamics by introducing an additional friction term, generically decreasing $n_s$ and increasing $r$. Moreover, the scalar and tensor power spectra at high frequencies are strongly enhanced, leading to a wide spectrum of potentially observable features. In particular the GW spectrum, experimentally accessible over a very wide range of frequencies, can provide valuable information on the key parameters of the scalar potential driving inflation.

\section*{Acknowledgments}
We acknowledge the financial support of the UnivEarthS Labex program at Sorbonne Paris Cit\'e (ANR-10-LABX-0023 and ANR-11-IDEX-0005-02).

\end{document}